\documentclass[10pt]{article}
\usepackage[OE]{express}

\begin{document}
\title{Carrier-envelope phase controlled isolated attosecond pulses in the
nm wavelength range, based on superradiant nonlinear Thomson-backscattering}

\author{Szabolcs Hack,\authormark{1} S\'{a}ndor Varr\'{o},\authormark{1,2} and Attila Czirj\'{a}k\authormark{1,3,*}}

\address{\authormark{1}ELI-ALPS, ELI-HU Non-Profit Ltd., H-6720 Szeged, Dugonics
t\'{e}r 13, Hungary\\
\authormark{2}Wigner Research Center for Physics, SZFI, H-1525 Budapest,
POBox 49, Hungary\\
\authormark{3}Department of Theoretical Physics, University of Szeged, H-6720
Szeged, Tisza L. krt. 84-86, Hungary}

\email{\authormark{*}czirjak@physx.u-szeged.hu} 



\begin{abstract}
A proposal for a novel source of isolated attosecond XUV -- soft X-ray
pulses with a well controlled carrier-envelope phase difference (CEP)
is presented in the framework of nonlinear Thomson-backscattering.
Based on the analytic solution of the Newton-Lorentz equations, the
motion of a relativistic electron is calculated explicitly, for head-on
collision with an intense fs laser pulse. By using the received formulae,
the collective spectrum and the corresponding temporal shape of the
radiation emitted by a mono-energetic electron bunch can be easily
computed. For certain suitable and realistic parameters, single-cycle
isolated pulses of ca. 20 as length are predicted in the XUV -- soft X-ray
spectral range, including the 2.33-4.37 nm water window. According
to our analysis, the generated almost linearly polarized beam is extremely
well collimated around the initial velocity of the electron bunch,
with considerable intensity and with its CEP locked to that of the
fs laser pulse. 
\end{abstract}


%
%

\bibliographystyle{osajnl}
\bibliography{forras_OE_3}

\section{Introduction}

Isolated attosecond XUV pulses allow us to investigate the real time
electron dynamics in atoms, molecules and solids experimentally \cite{Krausz_Ivanov}.
It is well known, that the carrier envelope phase difference (CEP)
of the femtosecond laser pulse, involved in most of these pioneering
experiments, affects various processes \cite{Krausz_CEP_control,Goulemakis_2010,kruger_nanotip}
in atomic or molecular systems on this time scale. Recently, it was
predicted that it is also crucial to control the CEP of the attosecond
pulses in these pump-probe experiments \cite{Peng_XUV_CEP_2007,Sansone_XUV_CEP_2013,Tibai_PRL_2014}.

Currently, the established way to generate attosecond XUV pulses is
based on high order harmonic generation in noble gas samples \cite{sansone_2010},
which has its limitations both in pulse length and intensity. In this
contribution, we are going to show that nonlinear Thomson-backscattering
provides a very promising method to generate isolated attosecond pulses
in the XUV -- soft X-ray spectral range with remarkable pulse properties.

Nonlinear Thomson-backscattering of a high intensity laser pulse on
a bunch of relativistic electrons \cite{esarey_1993} has long been
used as a source of X- and gamma-ray radiation \cite{Lee_attopulse_Thomson,Yan_2017_NatPhot},
usually with an emphasis on monochromatic features \cite{Sarri_2014_PRL,Khrennikov_2015_PRL}
or producing pulses of ps or fs length \cite{schoenlein_1996_science,Phuoc_2012_NatPhot}.
For a review see e.g. \cite{Corde_review_2013} and references therein.
To our best knowledge, results on attosecond (and even shorter) pulses
or pulse trains based on this process were published only in the hard
X- and gamma-ray spectral range \cite{Chung_2009_OptExp,Luo_2014_OptExp,Li_2015_PRL}.

The generation of electron bunches suitable for nonlinear Thomson-backscattering
(i.e fs and sub-fs pulse length, low emittance, sufficient density
and energy, small enough energy-spread) was promoted by pioneering
experiments \cite{Toth_Csaba_2004,sears_attobunch_experiment} and
enlightening simulation results \cite{Naumova_Mourou_atto_e_bunch_2004}
over the past two decades \cite{esarey_2009,Corde_review_2013}. More
recent developments include the utilization of velocity bunching to
generate an electron bunch with pC charge in the MeV energy range
\cite{Luo_2014_OptExp}, recently with already sub-10 fs pulse length \cite{Maxson_2017,Zhu_2016},
and a work on bunch compressing \cite{Sell_bunch_compressing_2014}
predicting electron bunches of 2 as duration and 5.2 MeV energy.

In this paper, based on our earlier works \cite{varro_thomson,sajat_NIMB},
we investigate in detail the radiation of a realistic attobunch of
electrons due to a near infrared (NIR) fs laser pulse in the $10^{18}-10^{19}\ \mathrm{W/cm^{2}}$
intensity range. First, we explicitly give the analytic solution of
the Newton-Lorentz equations for an electron moving in a plane wave
for a laser pulse with sine-squared envelope and with an arbitrary
number of cycles and CEP. Using this result, we compute the radiation
emitted by a bunch of $N$ electrons, both in frequency and in time
domain, and analyze the temporal and spatial profile and the CEP dependence
of the resulting isolated attosecond pulse.

\section{Analytic solution of the electron's equation of motion}

We assume that the laser pulse propagates in the $z$ direction and
it is linearly polarized along the $x$ direction. First, we consider
one electron only, which moves initially in the $-z$ direction, i.e.
we investigate a head-on collision. We model the electric field of
the laser pulse, $\mathbf{E}=(E_{x},0,0)$, with the usual sine-squared
envelope: 
\begin{equation}
E_{x}\left(\varTheta\right)=E_{0}\sin^{2}\left(\frac{\omega_{L}\varTheta}{2n_{c}}\right)\cos\left(\omega_{L}\varTheta-\varphi_{0}\right),\label{eq:laserfield}
\end{equation}
where $E_{0}$ is the amplitude, $\omega_{L}$ is the angular frequency,
$n_{c}$ is the number of optical cycles in the pulse, $\varphi_{0}$
is the CEP and $\varTheta=t-\mathbf{n}_{L}\mathrm{\mathbf{r}}/c$
is the wave argument of the laser pulse at position $\mathrm{\mathbf{r}}$,
with $\mathbf{n}_{L}$ denoting the unit vector pointing in the propagation
direction. The Newton-Lorentz equations govern the motion of a relativistic
electron with charge $e$ and mass $m$ during its interaction with
the laser pulse as 
\begin{eqnarray}
m\frac{d\mathbf{u}}{d\tau} & = & \frac{e}{c}\left[u^{0}\mathbf{E}\left(\varTheta\right)+\mathbf{n}_{L}\left(\mathbf{u}\mathbf{E}\left(\varTheta\right)\right)-\mathbf{E}\left(\varTheta\right)\left(\mathbf{n}_{L}\mathbf{u}\right)\right]\label{eq:N-L_space_propertime}\\
m\frac{du^{0}}{d\tau} & = & \frac{e}{c}\mathbf{E}\cdot\mathbf{u},\label{eq:N-L_time_propertime}
\end{eqnarray}
where $\left(u^{0},\mathbf{u}\right)=\left(\gamma c,\gamma\mathbf{v}\right)$
is the four-velocity, $\gamma\equiv\left(1-\left|\mathbf{v}\right|^{2}/c^{2}\right)^{-1/2}$
is the Lorentz-factor and $d\tau=dt/\gamma$ is the proper time element
of the electron. In (\ref{eq:N-L_time_propertime}) we have made use
of the $\mathbf{B}=\mathbf{n}_{L}\times\mathbf{E}/c$, connecting
the magnetic induction and the electric field strength of a plane
wave. As it is well known, the equations of motion (\ref{eq:N-L_space_propertime})-(\ref{eq:N-L_time_propertime})
have a general analytic solution due to the following linear relation
between the proper time of the electron and wave argument \cite{Sengupta_1949,mcmillan_1950,Goldman_1964}:
\begin{equation}
u^{0}-u^{3}=\frac{d}{d\tau}\left(ct-z\right)=c\frac{d\varTheta}{d\tau}=\alpha c,\label{eq:alpha}
\end{equation}
where $\alpha=\gamma(1-v_{z}/c)$ is a dimensionless constant of motion
depending on the initial conditions of the electron only. We have
determined the solution of (\ref{eq:N-L_space_propertime}-\ref{eq:N-L_time_propertime})
for the pulse shape \eqref{eq:laserfield} explicitly, which reads
as

\begin{eqnarray}
x\left(\varTheta\right) & = & x\left(\varTheta_{0}\right)+V_{x_{0}}\left(\varTheta-\varTheta_{0}\right)+c\varOmega\left(\varTheta\right),\label{eq:x_theta}\\
y\left(\varTheta\right) & = & y\left(\varTheta_{0}\right)+V_{y_{0}}\left(\varTheta-\varTheta_{0}\right),\label{eq:y_theta}\\
z\left(\varTheta\right) & = & z\left(\varTheta_{0}\right)+\varLambda\left(\varTheta-\varTheta_{0}\right)+V_{x_{0}}\varOmega\left(\varTheta\right)+\varDelta\left(\varTheta\right).\label{eq:z_theta}
\end{eqnarray}
The $t\left(\varTheta\right)$ component has the same functional form
as the $z\left(\varTheta\right)$ according to equation \eqref{eq:alpha},
they differ in the initial conditions only. We introduced above the
following quantities, having the dimension of velocity: 
\begin{eqnarray}
V_{x_{0}} & = & \alpha^{-1}u^{1}\left(\varTheta_{0}\right)+cf\left(\varTheta_{0}\right),\\
V_{y_{0}} & = & \alpha^{-1}u^{2}\left(\varTheta_{0}\right),\\
V_{z_{0}} & = & \alpha^{-1}u^{3}\left(\varTheta_{0}\right)+g\left(\varTheta_{0}\right)+h\left(\varTheta_{0}\right)+l\left(\varTheta_{\text{0}}\right),
\end{eqnarray}
with

\begin{align}
f\left(\Theta\right) & =\sum_{j=-1}^{1}\left(-\frac{1}{2}\right)^{1+\left|j\right|}\frac{\nu\ n_{c}}{n_{c}+j}\sin\left(\frac{n_{c}+j}{n_{c}}\varTheta\omega_{L}+\varphi_{0}\right),\\
g\left(\varTheta\right) & =-\frac{c\nu^{2}}{2}\frac{n_{c}^{2}}{n_{c}^{2}-1}\sum_{j=1}^{2}\left(-\frac{1}{4}\right)^{j}\cos\left(j\frac{\varTheta\omega_{L}}{n_{c}}\right),\\
h\left(\varTheta\right) & =\frac{c\nu^{2}}{32}\frac{3n_{c}^{2}-2}{n_{c}^{2}-1}\cos\left(2\left(\varTheta\omega_{L}+\varphi_{0}\right)\right),\\
l\left(\varTheta\right) & =\frac{c\nu^{2}}{4}\sum_{k=1}^{2}\sum_{\begin{array}{c}
j=\{-1,1\}\end{array}}\left(-\frac{n_{c}}{4\left(n_{c}+j\right)}\right)\cos\left(\frac{2n_{c}+kj}{n_{c}}\varTheta\omega_{L}+2\varphi_{0}\right),
\end{align}
where $\nu=\left|e\right|E_{0}/mc\omega_{L}\alpha=a_{0}/\alpha$ is
the effective intensity parameter, and $a_{0}=8.5 \times 10^{-10}\lambda\left[\mathrm{\mu m}\right]\sqrt{I_{0}\left[W/\mathrm{cm}^{2}\right]}$
denotes the dimensionless vector potential (the usual intensity parameter).
The $\varOmega\left(\varTheta\right)$ is an oscillating function
defined as 
\begin{equation}
\varOmega\left(\varTheta\right)=-\intop_{\varTheta_{0}}^{\varTheta}f\left(\Theta'\right)d\varTheta',\label{eq:omega_theta}
\end{equation}
making the $c\varOmega\left(\varTheta\right)$ to be the dominating
term in $x(\varTheta)$. The $\varLambda$ is a constant depending
on the initial values only: 
\begin{equation}
\varLambda=V_{z_{0}}+V_{x_{0}}f\left(\Theta_{0}\right).
\end{equation}
In $z\left(\varTheta\right)$, the $\varLambda$ is the most dominant
term because $V_{z_{0}}$ is larger than all the other terms for a
relativistic electron moving in the $z$ direction. The $\varDelta\left(\varTheta\right)$
is the well-known trajectory with systematic drift caused by the classical
radiation pressure: 
\begin{align}
\varDelta\left(\varTheta\right) & =-\intop_{\varTheta_{0}}^{\varTheta}\left[g\left(\Theta'\right)+h\left(\Theta'\right)+l\left(\Theta'\right)\right]d\varTheta'.\label{eq:delta_theta}
\end{align}
Since $f,\ g,\ h,\ l$ are linear combinations of simple trigonometric
functions of $\varTheta$, the explicit formulae for $\varOmega$
and $\varDelta$ can be easily obtained.

Due to the use of the wave argument $\varTheta$, the specification
of the initial values for the solution (\ref{eq:x_theta}-\ref{eq:z_theta})
requires some attention \cite{varro_thomson,sajat_NIMB}. The interaction
of an electron with the laser pulse starts if $\varTheta=\varTheta_{0}$
and it ends if $\varTheta=\varTheta_{1}$, i.e. these are specified
on a light-like hyper-surface. This means, that one has to transform
the usual initial conditions, which are valid in a lab-frame (i.e.
on a space-like hyper-surface), to the light-like hyper-surface. Ignoring
this important step leads to false peaks in the calculated spectrum,
as we demonstrated it in \cite{sajat_NIMB}.

\section{Emitted radiation spectra}

Now we proceed to evaluate the spectrum of radiation emitted by an
electron, moving according to the solution (\ref{eq:x_theta}-\ref{eq:z_theta}).
We specify an almost single-cycle \footnote{This terminology about the pulse length (FWHM) measured in the number
of cycles is commonly used in the laser physics community, although
the laser pulse has 3 optical cycles under the envelope function (see
inset on Fig. \ref{fig:more_e_cepfitting}). } sine laser pulse by setting $n_{c}=3$ and $\varphi_{0}=\pi/2$,
with a carrier wavelength of $\lambda_{L}=800$ nm and a dimensionless
vector potential of $a_{0}=1$, corresponding to a peak electric field
of ca. $4 \times 10^{12}$ V/m. The emitted radiation of an electron in
the far-field is given by the following formula \cite{jackson_angol}:
\begin{equation}
\mathbf{E_{1}}\left(\omega\right)=\frac{e}{c}\frac{e^{i\omega R_{0}/c}}{4\pi\varepsilon_{0}R_{0}}\intop_{-\infty}^{\infty}\frac{\mathbf{n}\times\left[\left(\mathbf{n}-\boldsymbol{\beta}\right)\times\boldsymbol{\dot{\beta}}\right]}{\left(1-\mathbf{n}\cdot\boldsymbol{\beta}\right)^{2}}e^{i\omega\left(t-\mathbf{n}\cdot\mathbf{r}\left(t\right)/c\right)}dt,\label{eq:Eomega_t}
\end{equation}
where $R_{0}$ is the distance of the observation point, $\mathrm{\mathbf{n}}$
is the unit vector pointing towards the observer, $\boldsymbol{\beta}=\mathbf{v}/c$
and $\boldsymbol{\dot{\beta}}$ are the normalized velocity and acceleration,
respectively. Here we note that in case of a charge interacting with
a fs laser pulse it is essential to use \eqref{eq:Eomega_t} which
includes also the end point terms that are usually neglected \cite{chen_2013}.

\begin{figure}
\begin{center}
\includegraphics[scale=0.37]{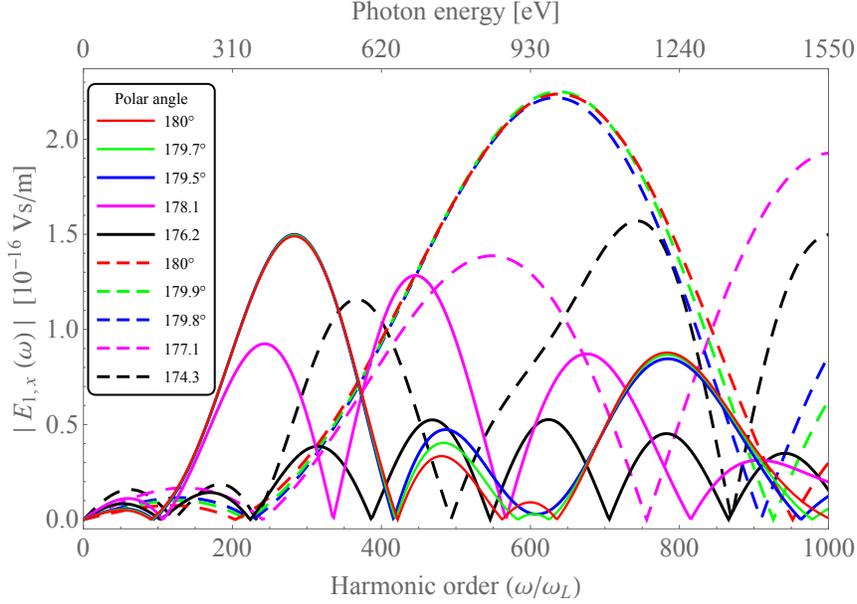}
\end{center}
\caption{\label{fig:spectrum} Nonlinear Thomson-backscattering spectra of
a single electron in head-on collision with a single-cycle laser pulse
of sine-squared envelope, Eq. (\ref{eq:Eomega_t}). We plot the spectra
of the dominant $x$ component of the electric field for $\gamma_{0}=10$
(solid lines) and for $\gamma_{0}=15$ (dashed lines), along the propagation
directions in the $x-z$ plane, specified by the polar angles in the
legend. ($E_{1,x}(\omega)$ is at least 1000 times larger than $E_{1,y}(\omega)$
or $E_{1,z}(\omega)$.) Other parameters: $\lambda_{L}=800$\ nm,
$n_{c}=3$, $a_{0}=1$, $R_{0}=2$\ m.}
\end{figure}

\begin{figure}
\begin{center}
\includegraphics[scale=0.37]{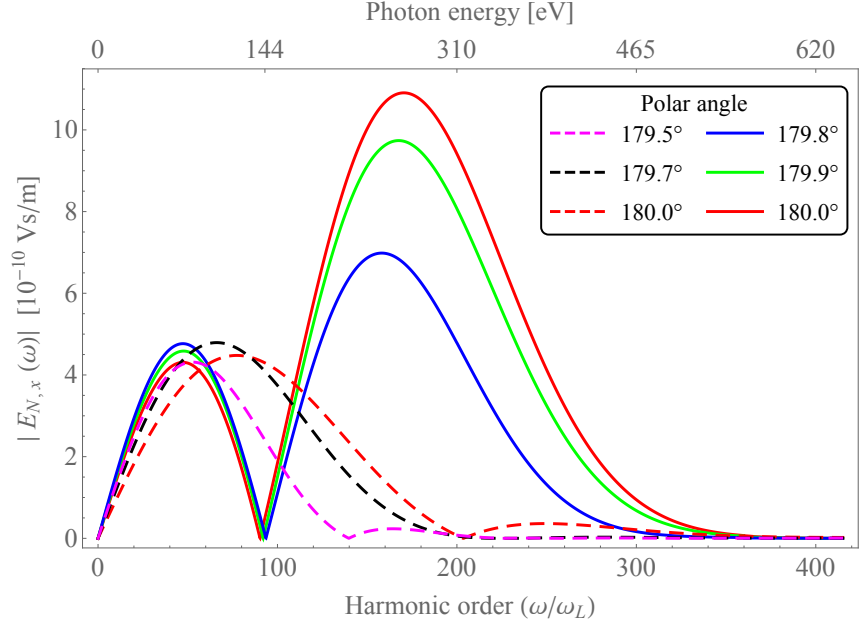}
\end{center}
\caption{\label{fig:more_e_spectrum} Nonlinear Thomson-backscattering spectra,
radiated collectively by an attobunch of $10^{8}$ electrons, having
an initial Lorentz factor of $\gamma_{0}=10$ (solid line) and $\gamma_{0}=15$
(dashed line). We plot the spectra of the dominant $x$ component
of the electric field, along the propagation directions in the $x-z$
plane specified by the polar angles in the legend. Other parameters
are the same as for Fig. \ref{fig:spectrum}.}
\end{figure}

By changing the integration variable from $t$ to $\varTheta$, we
can use the analytic trajectories (\ref{eq:x_theta}-\ref{eq:z_theta})
for calculating the emitted radiation. The resulting single electron
radiation spectrum is shown in Fig. \ref{fig:spectrum} for two selected
values of the initial Lorentz factor $\gamma_{0}$, along the directions
$\mathrm{\mathbf{n}}$ in the $x-z$ plane defined by the indicated
polar angles $\vartheta$ (i.e. very close to the direction of the
electron's initial velocity at $180\text{\textdegree}$). Two of these
polar angles were selected according to the usual $1/\gamma_{0}$
divergence of the radiation, while the other two polar angles, extremely
close to $180\text{\textdegree}$, are specified in accordance with
the collective spectra of Fig. \ref{fig:more_e_spectrum}.

The spectra and their angular dependence are similar for the two values
of $\gamma_{0}$, although for $\gamma_{0}=15$ the spectral peaks
are up-shifted and broadened compared to those for $\gamma_{0}=10$,
showing the strong influence of the initial relativistic velocity
of the electron on the spectrum \cite{Salamin_2003}. The nearly single-cycle
length of the NIR laser pulse causes further spectral broadening on
Fig. \ref{fig:spectrum}, which makes them more different form those
calculated earlier for the usual long laser pulses, especially when
approximated by a continuous wave laser field \cite{esarey_1993,Hartemann_1996,thomas_2010,rykovanov_2014}.

Based on these results, let us now consider the collective radiation
of an attobunch of electrons, which consists typically of $10^{5}-10^{8}$
electrons and has its longitudinal size $\ell$ in the 1-100 nm range.
In particular, we use electron attobunch parameters based on
the simulations of Naumova et. al. \cite{Naumova_Mourou_atto_e_bunch_2004} 
and on the predictions of Sell and K\"{a}rtner  \cite{Sell_bunch_compressing_2014}:
it consists of $N=10^{8}$ electrons with negligible energy spread,
its distribution is uniform with a size of 800\ nm ($=\lambda_{L}$)
in the transverse direction, while its distribution is Gaussian with
a size of 8\ nm (6 standard deviation) in the longitudinal direction. Other experimental
and simulation results, like e.g. \cite{Toth_Csaba_2004,sears_attobunch_experiment},
also suggest that these attobunch parameters are within reach experimentally
in the near future. These parameters, taking into account also the
high intensity and the few fs length of the laser pulse, justify to
treat this attobunch as an ideal electron bunch, i.e. we may safely
neglect its energy spread and transverse momentum, the radiation reaction
and the electron-electron interaction (the Coulomb-force between the
electrons is three orders of magnitude smaller than the Lorentz-force
due to the laser pulse for $a_0=1$).

Then we can generalize equation \eqref{eq:Eomega_t} to describe the
collectively emitted nonlinear Thomson-backscattered radiation of
$N$ electrons with the help of the coherence factor (sometimes called
also relativistic form factor) \cite{varro_thomson,Corde_review_2013}:
\begin{equation}
C_{N}\left(\omega\right)=\sum_{k=1}^{N}\exp\left[i\omega\left(t_{k}\left(\varTheta_{0}\right)-\frac{\mathbf{n}\cdot\mathbf{r}_{k}\left(\varTheta_{0}\right)}{c}\right)\right],\label{eq:FormFactor}
\end{equation}
which takes into account the effect of the different initial positions
of the electrons on the collectively emitted spectrum of $N$ electrons
as: 
\begin{equation}
\mathbf{E}_{N}\left(\omega\right)=C_{N}\left(\omega\right)\mathbf{E_{1}}\left(\omega\right).\label{eq:CollectiveSpectrum}
\end{equation}

\begin{figure}
\begin{center}
\includegraphics[scale=0.37]{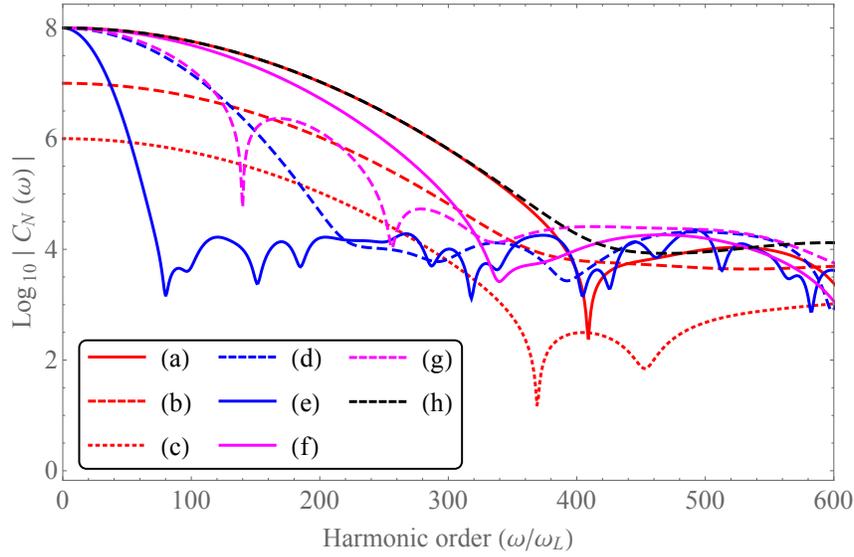} 
\end{center}
\caption{\label{fig:coherencefactor} Magnitude of the coherence factor \eqref{eq:FormFactor}
for a particular random realization of the ideal electron attobunch
described in the text. Parameters for curve (a): $N=10^{8}$, $\gamma_{0}=10$,
$\vartheta=180\text{\textdegree}$, $\ell=8\:\text{nm}$. Parameter
changes with respect to (a) for (b): $N=10^{7}$; (c): $N=10^{6}$;
(d): $\ell=15\:\text{nm}$; (e): $\ell=45\:\text{nm}$; (f): $\vartheta=179.8\text{\textdegree}$;
(g): $\vartheta=179.5\text{\textdegree}$; (h): $\gamma_{0}=15$.}
\end{figure}

The sensitive dependence of the coherence factor on certain parameters
influences the collective radiation in a non-trivial way, therefore
we examine first the magnitude of $C_{N}\left(\omega\right)$ in Fig.
\ref{fig:coherencefactor} on a logarithmic scale. For the attobunch
parameters specified above, the shape of the curve (a) is independent
of the particular set of individual electron coordinates at least
up to the 400th harmonics. Although it exhibits a slight fluctuation
above this value, but this does not influence the collective spectrum,
since its magnitude is negligible already. Comparison of curves (a-c)
clearly shows that the magnitude of the coherence factor scales linearly
with the number of electrons, predicting the possibility of a superradiant
collective emission. Note that the frequency range free of fluctuations
slightly decreases with decreasing $N$. Comparison of curves (a),
(d) and (e) shows that the frequency range of constructive coherent
superposition is decreased inversely proportionally with the increasing longitudinal
size of the attobunch. Comparison of curves (a), (f) and (g) shows
that slight changes in the direction of the radiation have a very
similar effect. However, curves (a) and (h) show, the the coherence
factor is not sensitive to the value of the initial Lorentz-factor
in this range.

Next we show the spectra of the collective radiation, computed on
the basis of Eq. (\ref{eq:CollectiveSpectrum}), in Fig. \ref{fig:more_e_spectrum}
for $\gamma_{0}=10$ (solid lines) and $\gamma_{0}=15$ (dashed lines)
along the directions defined by the indicated polar angles in the
$x-z$ plane, in accordance with three of the spectra in Fig. \ref{fig:spectrum}.
Note that a considerable portion of this radiation is in the 2.33-4.37
nm (i.e. 283.7-532.1 eV) water window (especially for $\gamma_{0}=10$)
which may provide an important possibility in the experimental study
of organic molecules in water environment \cite{Nisoli_ChemRev_2017}.

In agreement with the sensitive dependence of $C_{N}(\omega)$ on
the polar angle, the attobunch creates its collective radiation in
a superradiant manner only in a narrow cone with an opening angle
of a few tenth degrees, which means a bright beam with an extremely
small divergence compared to the usual case of nonlinear Thomson-backscattering.
(We note that although the term superradiance was introduced in quantum
optics for a process which involves also an interaction between the
emitters mediated by the field \cite{Dicke_Superradiance}, here we
have independent emitters and we use the term superradiance only to
emphasize that the intensity of the emitted radiation depends quadratically
on the number of electrons in the bunch \cite{Schott_1912}.) In case
of $\gamma_{0}=15$, unlike the expectation, the divergence of the
emitted radiation does not decrease further but it is somewhat broader
than for $\gamma_{0}=10$. Note also that for $\gamma_{0}=15$ the
maximum of $\left|\mathbf{E}\left(\omega\right)\right|$ is not in
the direction of the initial velocity of the electron bunch, as for
$\gamma_{0}=10$.

\begin{figure}
\begin{center}
\includegraphics[scale=0.37]{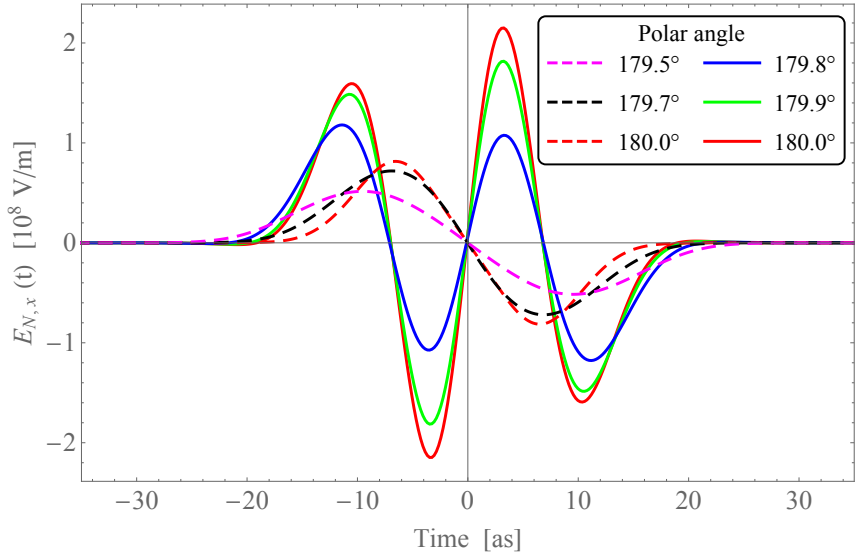} 
\end{center}
\caption{\label{fig:more_e_pulseandspot} Temporal pulse shapes of the isolated
attosecond pulses obtained by nonlinear Thomson-backscattering, computed
from the spectra on Fig. \ref{fig:more_e_spectrum}, corresponding
to $\gamma_{0}=10$ (solid line) and $\gamma_{0}=15$ (dashed line).
We plot the dominant $x$ component of the electric field. Parameters
are the same as for Fig. \ref{fig:spectrum}.}
\end{figure}

\section{Properties of the emitted isolated attosecond pulses}

We show the temporal pulse shapes of the collective radiation in Fig.
\ref{fig:more_e_pulseandspot}, based on the inverse Fourier-transform
of the corresponding collective spectra of Fig. \ref{fig:more_e_spectrum}.
Remarkably, we have an isolated attosecond pulse for both of the values
of $\gamma_{0}$, however, with different pulse shapes. Note also,
that this pulse shape does not change considerably along the radiation
directions with different polar angles and it is independent of the
azimuthal angle, i.e. the pulse shapes are ca. the same within the
beam spot.

For $\gamma_{0}=10$, the pulse has only two oscillations and its
length at FWHM is 22.5 as. In a distance of $R_{0}=2$\ m from the
interaction region, the peak intensity is $6.14\times 10^{9}\:\mathrm{W/cm^{2}}$
and the average intensity is $1.81\times 10^{9}\:\mathrm{W/cm^{2}}$, giving
a pulse energy of $60.86$ nJ. For $\gamma_{0}=15$, the pulse has only
one single oscillation and its length at FWHM is 19.2 as. For $R_{0}=2$\ m,
the peak intensity is $9.68\times 10^{8}\:\mathrm{W/cm^{2}}$ and the average
intensity is $5.55\times 10^{8}\:\mathrm{W/cm^{2}}$, giving a pulse energy
of $18.68$ nJ.

Regarding the polarization of the pulse, the $x$ component of the
electric field is at least 3 orders of magnitude larger than its $z$
component. For non-zero values of the azimuthal angle, the radiation
has also a $y$ component which is similar in magnitude to the $z$
component. However, $E_{N,y}(t)$ is not in phase with the dominant
$x$ component which makes the polarization of the pulse non-trivial
around the nodes of the $x$ component. Nevertheless, this can be
easily corrected for in an experiment if one wishes to have perfect
linear polarization.

\begin{figure}
\begin{center}
\includegraphics[scale=0.37]{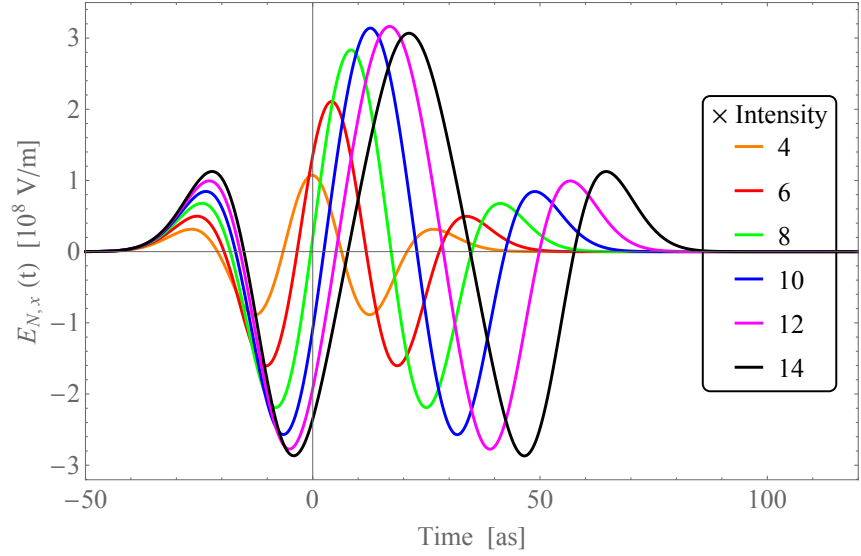}
\end{center}
\caption{\label{fig:more_e_intensity_scan} Temporal pulse shapes of the isolated
attosecond pulses, obtained by nonlinear Thomson-backscattering at
$\vartheta=180\text{\textdegree}$, in case of the indicated values of $a_{0}^2$
for the NIR cosine-type ($\varphi_{0}=0$) laser pulse. 
The electron bunch parameters correspond to curve (d) in Fig. 
\ref{fig:coherencefactor}.}
\end{figure}

The above values of pulse energy and intensity are already
high enough for state of the art pump-probe experiments. However,
the quadratic dependence of these quantities on $N$ in the superradiant
parameter range may provide several orders of magnitude larger values,
since 1 or 2 orders of magnitude increase in the number of electrons
in the attobunch seams to be feasible. 
Another way of increasing the pulse energy and intensity is to increase the intensity of the NIR pulse. 
We plot the temporal shapes of the resulting attosecond pulses along the direction of 
$\vartheta=180\text{\textdegree}$  in Fig. \ref{fig:more_e_intensity_scan}, 
corresponding to $a_{0}^2$ values in the range of 4 to 12. 
Here we assume a cosine-type NIR pulse and a longer electron attobunch with the parameters corresponding to 
curve (d) in Fig. \ref{fig:coherencefactor}. (Note also, that this longer electron attobunch generates lower intensity pulses than the one used in the case of Fig. \ref{fig:more_e_pulseandspot}.) 
The plots of Fig. \ref{fig:more_e_intensity_scan} show that the intensity of the attosecond pulse increases nonlinearly with increasing NIR intensity up to a  
saturation intensity, while the pulse length increases only very moderately.
E.g. for $a_{0}^2=10$, the pulse length is still not more than $45$ as, 
but the peak intensity is already $1.31\times 10^{10}\:\mathrm{W/cm^{2}}$ 
and the average intensity is $5.54\times 10^{9}\:\mathrm{W/cm^{2}}$, 
giving a pulse energy of $381.69$ nJ.
These results suggest that there is an optimal NIR laser intensity 
for a given set of bunch parameters, which already yields the highest possible intensity of the attosecond pulse while its pulse length is still the shortest possible
at that intensity. 

\begin{figure}
\begin{center}
\includegraphics[scale=0.37]{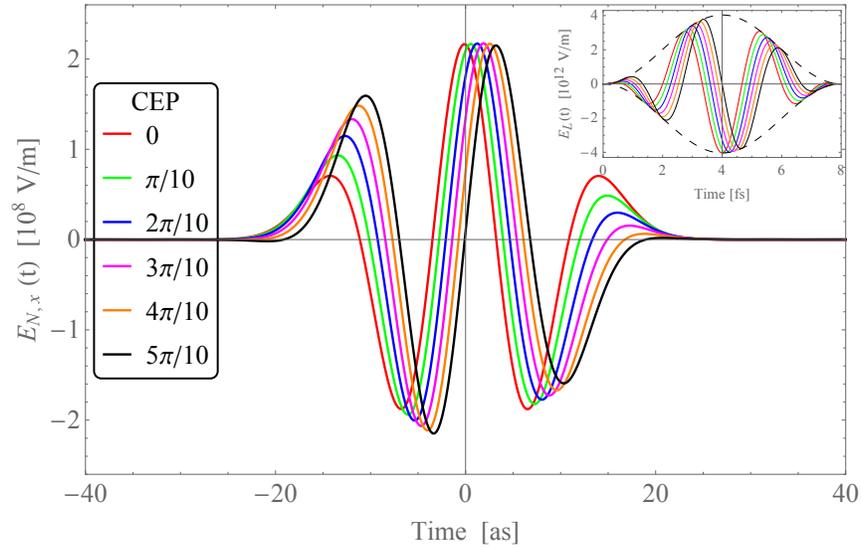}
\end{center}
\caption{\label{fig:more_e_cepfitting} Temporal pulse shapes of the isolated
attosecond pulses, obtained by nonlinear Thomson-backscattering at
$\vartheta=180\text{\textdegree}$, for different values of the CEP
of NIR laser pulse given in the legend. The inset shows the pulse
shapes of the incoming NIR pulses of different CEP with the corresponding
colors. The $\gamma_{0}=10$, other parameters are the same as for
Fig. \ref{fig:spectrum}.}
\end{figure}

Finally, we discuss the CEP dependence of the emitted attosecond pulses
on the CEP of the single-cycle NIR laser pulse. Since this latter
is an independent parameter in the solutions (\ref{eq:x_theta}-\ref{eq:z_theta}),
it is straightforward to calculate the pulse shapes emitted by the
attobunch for any value of the CEP of the NIR laser pulse. We show
the results of this investigation in Fig. \ref{fig:more_e_cepfitting}:
the CEP of the attosecond pulse perfectly follows the CEP of the NIR
laser pulse with a phase difference of $\pi$. This very simple relationship
makes the CEP of these attosecond pulses easily controllable through
the CEP of the NIR laser pulse, which is expected to have growing
importance in attosecond pump and probe experiments.

\section{Summary and conclusions}

As a summary, we investigated the nonlinear Thomson-backscattering
of a NIR laser pulse on an (ideally treated) relativistic electron
bunch, based on an explicit analytic solution of the Newton-Lorentz
equations which is valid for a frequently used laser pulse shape family.
Our result show that an attobunch of $10^{8}$ electrons having 5.2
MeV energy could produce an isolated XUV -- soft X-ray pulse of 22.5
as length and 60.86 nJ energy, and with its CEP locked to the CEP of the
NIR laser pulse. Based on the analysis of the coherence factor, we
identified the important parameters of this superradiant process which
may further enhance the pulse intensity by orders of magnitude. We
hope that these results promote further theoretical and experimental
research on XUV -- soft X-ray pulse sources based on Thomson-backscattering,
and on the generation of electron attobunches.

\section*{Funding}
The project has been supported by the European Union, co-financed by the European Social Fund, EFOP-3.6.2-16-2017-00005.
Partial support by the
ELI-ALPS project is also acknowledged. The ELI-ALPS project (GOP-1.1.1-12/B-2012-000,
GINOP-2.3.6-15-2015-00001) is supported by the European Union and
co-financed by the European Regional Development Fund.

\section*{Acknowledgments}
The authors thank M.\ G.\ Benedict, P.\ F\"{o}ldi, Zs.\ L\'{e}cz, D.\ Papp,
 Cs.\ T\'{o}th and Gy.\ T\'{o}th for stimulating discussions. 

\end{document}